\documentclass[pre,twocolumn,10pt,aps]{revtex4-1}

 \usepackage{bm}
 \usepackage{verbatim}
 \usepackage{amsmath}
 \usepackage{amssymb}
 \usepackage{amsthm}
 \usepackage{latexsym}
 \usepackage{amsfonts}
 \usepackage{epsfig}
 \usepackage{epstopdf}
 \usepackage{wasysym} 
 \usepackage{subfigure}
 \usepackage{color}
 \definecolor{darkblue}{rgb}{0,0,.5}
 \usepackage[linktocpage, colorlinks=true, linkcolor=darkblue, citecolor=darkblue]{hyperref}
 \usepackage[all]{hypcap}
 \usepackage[makeroom]{cancel}

\newcommand{\C}[1]{{\cal{#1}}}
\newcommand{\bb}[1]{\textbf{#1}}
\newcommand{\lr}[1]{{\left\langle {#1}\right\rangle}}

\begin{document}

\title{Measurability of nonequilibrium thermodynamics in terms of the Hamiltonian of mean force}

\author{Philipp Strasberg$^1$}
\author{Massimiliano Esposito$^2$}
\affiliation{$^1$F\'isica Te\`orica: Informaci\'o i Fen\`omens Qu\`antics, Departament de F\'isica, Universitat Aut\`onoma de Barcelona, 08193 Bellaterra (Barcelona), Spain}
\affiliation{$^2$Complex Systems and Statistical Mechanics, Departement of Physics and Materials Science, University of Luxembourg, L-1511 Luxembourg, Luxembourg}

\date{\today}

\begin{abstract}
The nonequilibrium thermodynamics of an open (classical or quantum) system in strong contact with a single heat bath can 
be conveniently described in terms of the Hamiltonian of mean force. However, the conventional formulation is limited by 
the necessity to measure differences in equilibrium properties of the system-bath composite. We make use of the freedom 
involved in defining thermodynamic quantities, which leaves the thermodynamics unchanged, to show that the 
Hamiltonian of mean force can be inferred from measurements on the system alone, up to that irrelevant freedom. 
In doing so, we refute a key criticism expressed in Phys.~Rev.~E \bb{94}, 022143 and arXiv:1911.11660. 
We also discuss the remaining part of the criticism. 
\end{abstract}

\maketitle

\newtheorem{lemma}{Lemma}[section]

\section{Introductory Review}
\label{sec intro}

We start by reviewing recent progress in nonequilibrium strong coupling thermodynamics based on the 
Hamiltonian of mean 
force~\cite{JarzynskiJSM2004, CampisiTalknerHaenggiPRL2009, CampisiTalknerHaenggiJPA2009, HiltEtAlPRE2011, 
PucciEspositoPelitiJSM2013, SeifertPRL2016, StrasbergEtAlNJP2016, PhilbinAndersJPA2016, JarzynskiPRX2017, 
MillerAndersPRE2017, StrasbergEspositoPRE2017, AurellEnt2017, PerarnauLlobetEtAlPRL2018, AurellPRE2018, 
HsiangHuEntropy2018, MillerAndersNC2018, StrasbergEspositoPRE2019, StrasbergPRL2019, 
HerpichShayanfardEspositoArXiv2019, StrasbergQuantum2020} in a unified way. The goal is to find a consistent 
thermodynamic description for a system, which 
can be driven far away from equilibrium and which exchanges energy with a single arbitrary strongly coupled bath. If it can 
also exchange particles with the bath, similar constructions were independently proposed in Refs.~\cite{BruchEtAlPRB2016, 
OchoaBruchNitzanPRB2016, HaughianEspositoSchmidtPRB2018}. We here focus only on the exchange of energy. 

For this purpose we consider open systems specified by an arbitrary global Hamiltonian of the form 
$H_{SB}(\lambda_t) = H_S(\lambda_t) + V_{SB} + H_B$. Here, $H_S(\lambda_t)$ is the system Hamiltonian, which can 
depend on an external driving protocol $\lambda_t$ (e.g., a changing electromagnetic field), $H_B$ is the bare bath  Hamiltonian, and $V_{SB}$ describes the system-bath interaction. Note that a time-dependent interaction 
$V_{SB}(\lambda_t)$ can be considered within the thermodynamic framework based on the Hamiltonian of mean 
force~\cite{StrasbergEspositoPRE2017, AurellEnt2017, PerarnauLlobetEtAlPRL2018, AurellPRE2018, StrasbergPRL2019}, but 
for ease of presentation we do not include this possibility here. In this paper, we use a quantum mechanical notation 
for convenience. If a result is only valid for classical systems, we explicitly emphasize it. 

For the moment we consider the time $t$ to be fixed. 
If the global state of the system-bath composite is an equilibrium 
canonical state at inverse temperature $\beta = T^{-1}$ ($k_B\equiv 1$), it is given by 
$\pi_{SB}(\lambda_t) \equiv e^{-\beta H_{SB}(\lambda_t)}/\C Z_{SB}(\lambda_t)$ 
with the partition function $\C Z_{SB}(\lambda_t) = \mbox{tr}\{e^{-\beta H_{SB}(\lambda_t)}\}$. 
In that case the corresponding reduced equilibrium state of the system is given by 
\begin{equation}\label{eq reduced state}
 \pi^*_S(\lambda_t) \equiv \mbox{tr}_B\{\pi_{SB}(\lambda_t)\}
\end{equation}
which is in general not of the canonical Gibbs form due to the non-negligible coupling $V_{SB}$, 
i.e., $\pi_S^*(\lambda_t) \neq e^{-\beta H_S(\lambda_t)}/\C Z_S(\lambda_t)$. 
However, it can be always written in that canonical form with an \emph{effective} Hamiltonian, which is known 
as the Hamiltonian of mean force (HMF)~\cite{OnsagerCR1933, KirkwoodJCP1935} and which equals 
$-T\ln\pi_S^*(\lambda_t)$ up to an additive constant. A common and convenient choice for that constant is 
fixed using $H_S^*(\lambda_t) = -T\ln[\C Z_S^*(\lambda_t)\pi_S^*(\lambda_t)]$ with~\cite{RouxSimonsonBC1999, 
GelinThossPRE2009} 
\begin{equation}\label{eq HMF}
 \C Z^*_S(\lambda_t) \equiv \frac{\C Z_{SB}(\lambda_t)}{\C Z_B}.
\end{equation}
Here, $\C Z_B \equiv \mbox{tr}_B\{e^{-\beta H_B}\}$ is the partition function of the bath alone. 
One should note that the HMF $H^*_S(\lambda_t) = H^*_S(\lambda_t,\beta)$ depends on the inverse temperature. 

In the following we recapitulate some essential elements of the nonequilibrium thermodynamics based on the HMF. 
For ease of presentation we assume that the initial system-bath state is described by the global equilibrium state
$\pi_{SB}(\lambda_0)$, where we set the initial time to be $t=0$. Note that this initial state is different from the class 
of decorrelated initial states $\rho_{SB}(0) = \rho_S(0)\otimes\pi_B$, which is conventionally considered in the theory of 
open quantum systems~\cite{BreuerPetruccioneBook2002} and requires a thermodynamic treatment \emph{not} captured by the 
HMF~\cite{AndrieuxEtAlNJP2009, EspositoLindenbergVandenBroeckNJP2010, TakaraHasegawaDriebePLA2010, ReebWolfNJP2014, 
PtaszynskiEspositoPRL2019, RivasArXiv2019, StrasbergWinterArXiv}. The present framework is therefore 
particularly designed to treat initially correlated (and perhaps even entangled) system-bath states. Note that, 
classically, a larger class of correlated initial system states can be treated provided that the bath is initially in 
a conditional equilibrium state~\cite{SeifertPRL2016}, but this result has no straightforward analog in the quantum 
regime~\cite{StrasbergEspositoPRE2019}. However, if one pays attention to the fact that the state preparation itself has 
a thermodynamic cost, then any initial system state can be also treated within the HMF framework of strong coupling 
thermodynamics~\cite{StrasbergPRL2019}. Finally, a single framework combining both, correlated and decorrelated initial 
states, was proposed in Ref.~\cite{RivasArXiv2019}. 

Although we assume to start in equilibrium, we allow the driving protocol $\lambda_t$ to vary arbitrarily 
in time. This implies that the system-bath state at a later time $t>0$ is no longer in equilibrium, i.e., 
$\rho_{SB}(t) \neq \pi_{SB}(\lambda_t)$ in general. The mechanical work done on the system is identified 
as usual as 
\begin{equation}\label{eq work}
 W(t) \equiv \int_0^t ds \mbox{tr}_S\left\{\frac{\partial H_S(\lambda_s)}{\partial s}\rho_S(s)\right\},
\end{equation}
where $\rho_S(t) = \mbox{tr}_B\{\rho_{SB}(t)\}$ is the reduced state of the system at time $t$.
The second law of nonequilibrium thermodynamics was found to be (for classical dynamics this was first derived in 
Ref.~\cite{SeifertPRL2016} and for quantum dynamics in Ref.~\cite{StrasbergEspositoPRE2019}) 
\begin{equation}\label{eq 2nd law}
 \Sigma(t) \equiv \beta[W(t) - \Delta F_S^*(t)] \ge 0
\end{equation}
with $\Delta F_S^*(t) \equiv F_S^*(t) - F_S^*(0)$. Here, the generalization of the \emph{nonequilibrium} free energy to 
the strong coupling regime is defined as 
\begin{equation}\label{eq free energy}
 F_S^*(t) \equiv \mbox{tr}_S\{H_S^*(\lambda_t)\rho_S(t)\} + T\mbox{tr}_S\{\rho_S(t)\ln\rho_S(t)\}.
\end{equation}
Furthermore, $\Sigma(t)$ is known as the entropy production and thus, Eq.~(\ref{eq 2nd law}) takes on the familiar 
form of phenomenological nonequilibrium thermodynamics~\cite{KondepudiPrigogineBook2007} identifying the free 
energy~(\ref{eq free energy}) as the quantity, which gets minimized at equilibrium. 
The second law~(\ref{eq 2nd law}) can be also expressed in terms of the relative entropy 
$D[\rho\|\sigma] \equiv \mbox{tr}\{\rho(\ln\rho-\ln\sigma)\}$ as~\cite{StrasbergEspositoPRE2019} 
\begin{equation}
 \Sigma(t) = D[\rho_{SB}(t)\|\pi_{SB}(\lambda_t)] - D[\rho_S(t)\|\pi^*_S(\lambda_t)],
\end{equation}
from which the nonnegativity of $\Sigma(t)$ follows. Furthermore, if the dynamics are classical, 
it also holds that~\cite{StrasbergEspositoPRE2019} 
\begin{equation}\label{eq EP rel ent}
 \Sigma(t) = -\int_0^t ds \left.\frac{\partial}{\partial s}\right|_{\lambda_s} D[\rho_S(s)\|\pi_S^*(\lambda_s)].
\end{equation}
Here, the derivative is taken with respect to a fixed $\lambda_s$. Finally, notice that the nonequilibrium 
free energy can be linked to the equilibrium free energy, denoted with a caligraphic letter 
$\C F_S^*(\lambda_t) = -T\ln\C Z_S^*(\lambda_t)$, via the relation 
\begin{equation}
 F_S^*(t) - \C F_S^*(\lambda_t) = T D[\rho_S(t)\|\pi_S^*(\lambda_t)] \ge 0.
\end{equation}
Due to the nonnegativity of relative entropy and since we assumed to start in equilibrium, 
Eq.~(\ref{eq 2nd law}) implies the \emph{weaker} inequality 
\begin{equation}\label{eq W diss}
 W_\text{diss}(t) \equiv W(t) - \Delta\C F_S^*(\lambda_t) \ge 0.
\end{equation}
In this context $W_\text{diss}(t)$ is known as the ``dissipated work'' and Eq.~(\ref{eq W diss}) was first derived 
for classical dynamics in Ref.~\cite{JarzynskiJSM2004} and for quantum dynamics in 
Ref.~\cite{CampisiTalknerHaenggiPRL2009}. 

Remarkably, Eq.~(\ref{eq W diss}) can be extended to a fluctuation theorem~\cite{JarzynskiJSM2004, 
CampisiTalknerHaenggiPRL2009}
\begin{equation}\label{eq Jarzynski}
 \lr{e^{-\beta w(t)}} = e^{-\beta\Delta\C F_S^*(\lambda_t)}.
\end{equation}
Classically, $\lr{\dots}$ denotes an ensemble average over many trajectories and $w(t)$ is the stochastic work, 
which follows from evaluating Eq.~(\ref{eq work}) along a single trajectory, see Ref.~\cite{JarzynskiJSM2004} for 
details. Quantum mechanically, Eq.~(\ref{eq Jarzynski}) can be derived using the so-called 
``two-point-projective-energy-measurement scheme'' (TPPEMS) (see Refs.~\cite{EspositoHarbolaMukamelRMP2009, 
CampisiHaenggiTalknerRMP2011} for reviews). Furthermore, for classical dynamics also Eq.~(\ref{eq 2nd law}) 
can be extended to a fluctuation theorem~\cite{SeifertPRL2016}: 
\begin{equation}\label{eq fluctuation theorem}
 \lr{e^{-\beta[w(t) - \Delta f_S^*(t)]}} = 1.
\end{equation}
Here, $f_S^*(t)$ is the stochastic nonequilibrium free energy, see Appendix~\ref{sec app} for more details and a proof 
of Eq.~(\ref{eq fluctuation theorem}). The two fluctuation theorems~(\ref{eq Jarzynski}) 
and~(\ref{eq fluctuation theorem}) need to be distinguished in general. 
If the dynamics are such that for a fixed control parameter $\lambda_t$ the final nonequilibrium state $\rho_S(t)$ 
relaxes back to the equilibrium state (for instance, when the global system is weakly coupled to an ideal 
superbath), then Eq.~(\ref{eq fluctuation theorem}) implies Eq.~(\ref{eq Jarzynski}). Interestingly, a corresponding 
quantum version of Eq.~(\ref{eq fluctuation theorem}) is not known to exist for general open system dynamics. 

We now turn to the definition of internal energy, heat and system entropy. We emphasize, however, that the second 
law~(\ref{eq 2nd law}), together with the definition of work, Eq.~(\ref{eq work}), is sufficient to characterize the 
set of allowed state transformations and the overall dissipation of the process. Indeed, it is clear from the 
basic definition of the nonequilibrium free energy, 
\begin{equation}\label{eq free energy general}
 F_S^*(t) = \tilde U_S(t) - T\tilde S_S(t),
\end{equation}
that there are \emph{a priori} many options to define an internal energy $\tilde U_S(t)$ (which fixes the definition 
of heat via the first law) or a thermodynamic entropy $\tilde S_S(t)$ of the system (which fixes the 
definition of heat via the second law), without having any impact on the second law. Furthermore, all that matters 
for the second law is the change in nonequilibrium free energy, which leaves us with a further freedom since 
Eq.~(\ref{eq free energy}) is only fixed up to an irrelevant constant value with respect to a standard reference state. 
We review two convenient choices. 

One choice, which was used in Refs.~\cite{SeifertPRL2016, MillerAndersPRE2017, AurellEnt2017, AurellPRE2018, 
StrasbergEspositoPRE2019, StrasbergPRL2019, StrasbergQuantum2020} to construct a framework of nonequilibrium 
thermodynamics, identifies 
\begin{align}
 \tilde U_S(t) &\equiv U_S^*(t) \label{eq U Seifert} \\
 &=	\mbox{tr}_S\left\{\rho_S(t)\left[H_S^*(\lambda_t) + \beta\partial_\beta H_S^*(\lambda_t)\right]\right\}, \nonumber \\
 \tilde S_S(t) &\equiv S_S^*(t)	\label{eq S Seifert} \\
 &=	\mbox{tr}_S\left\{\rho_S(t)\left[-\ln\rho_S(t) + \beta^2\partial_\beta H_S^*(\lambda_t)\right]\right\}, \nonumber
\end{align}
which requires to evaluate the partial derivative $\partial_\beta H_S^*(\lambda_t)$. 
Furthermore, starting with $\C F_S^*(\lambda_t) = -T\ln\C Z_S^*(\lambda_t)$, a 
straightforward calculation reveals that~\footnote{Quantum mechanically, this calculation is not completely 
straightforward~\cite{HsiangHuEntropy2018}. } 
\begin{equation}\label{eq equilibrium thermo relation}
 \C U_S^*(\lambda_t) = \partial_\beta[\beta\C F_S^*(\lambda_t)], ~~~ 
 \C S_S^*(\lambda_t) = \beta^2\partial_\beta\C F_S^*(\lambda_t).
\end{equation}
Here, $\C U_S^*(\lambda_t)$ and $\C S_S^*(\lambda_t)$ are the equilibrium counterparts of $U_S^*(t)$ and $S_S^*(t)$ 
obtained by replacing $\rho_S(t)$ with $\pi_S^*(\lambda_t)$. 
Equation~(\ref{eq equilibrium thermo relation}) looks familiar from equilibrium statistical mechanics if one replaces 
$\C X^*_S(\lambda_t)$ by $\C X_S(\lambda_t)$, where $\C X$ is used to denote $\C F, \C U$ or $\C S$. Furthermore, it 
follows from $\C Z_S^*(\lambda_t) = \C Z_{SB}(\lambda_t)/\C Z_B$ that a certain additivity property holds: 
\begin{equation}\label{eq additivity}
 \C X_S^*(\lambda_t) = \C X_{SB}(\lambda_t) - \C X_B.
\end{equation}
This implies, e.g., that the equilibrium system internal energy plus the equilibrium internal energy 
of the bare, unperturbed bath is equal to the global internal energy of the system-bath composite. 
The energy and entropy of the system, however, remain in general not additive. Indeed, if the system 
$S$ is split into two subsystems, $S = X \otimes Y$, and if one follows the same logic as above 
by assigning $\C X_X^*(\lambda_t) \equiv \C X_{XYB}(\lambda_t) - \C X_{YB}$ and 
$\C X_Y^*(\lambda_t) \equiv \C X_{XYB}(\lambda_t) - \C X_{XB}$ to $X$ and $Y$, respectively, 
it no longer holds true that $\C X^*_X + \C X^*_Y + \C X_B = \C X_{XYB}$, i.e., 
$\C X^*_X + \C X^*_Y \neq \C X^*_{XY}$ in general. 

Another choice arises if one is only interested in the coarse-grained thermodynamics of an extended system, 
$S' = S \otimes R$, which by incorporating part of the bath, $R$, can be treated as weakly coupled to the remaining part 
of the bath. This strategy, which is based on tools from Refs.~\cite{SeifertEPJE2011, EspositoPRE2012, BoCelaniJSM2014}, 
can be used to show that for classical dynamics the following definition emerges 
naturally~\cite{StrasbergEspositoPRE2017}: 
\begin{equation}
 \tilde F_S(t) \equiv F_S^\text{CG}(t) = F_S^*(t) + \C F_R.
\end{equation}
Here, ``CG'' stands for coarse-graining and $\C F_R$ is the equilibrium free energy of the part of the bath 
that was incorporated in the extended system which obviously has no impact on the change in system nonequilibrium 
free energy as $\Delta F_S^\text{CG}(t) = \Delta F_S^*(t)$. 
Furthermore, relations formally identical to Eqs.~(\ref{eq equilibrium thermo relation}) and~(\ref{eq additivity}), 
but in each case with a redefined equilibrium value, can be also derived. 
A crucial observation made in Ref.~\cite{StrasbergEspositoPRE2017} was that the so-defined thermodynamic quantities 
$F_S^\text{CG}(t)$, $U_S^\text{CG}(t)$ and $S_S^\text{CG}(t)$ capture the \emph{full} nonequilibrium thermodynamics 
of the weakly coupled open system $S' = S\otimes R$ in the limit where the remaining degrees of freedom $R$ are fast 
and can be adiabatically eliminated, i.e., whenever they can be approximated to be in a conditional equilibrium state. 
Even beyond that limit, so-called Markovian embedding strategies can be used to study the thermodynamics of 
strongly coupled open quantum systems~\cite{StrasbergEtAlNJP2016, NewmanMintertNazirPRE2017, SchallerEtAlPRB2018, 
StrasbergEtAlPRB2018, RestrepoEtAlNJP2018, SchallerNazirBook2018, RestrepoEtAlPRB2019, CorreaEtAlJCP2019}. 

We remark that all the results mention so far are powerful because they are exact mathematical identities that 
hold for any arbitrary system-bath Hamiltonian dynamics, and in particular any bath size.

\section{Local measurability of the Hamiltonian of mean force}

In the previous section we have reviewed a thermodynamic framework, where all thermodynamic quantities can be 
evaluated based solely on knowledge of the reduced system state $\rho_S(t)$. From the point of open quantum system 
theory~\cite{BreuerPetruccioneBook2002} this makes it an appealing theoretical framework. Also experimentally, while 
still challenging, quantum state tomography of $\rho_S(t)$ has been already demonstrated for many technologically 
relevant platforms. Classically, one can directly use stochastic trajectories to evaluate the corresponding stochastic 
thermodynamic quantities. 

However, there is one caveat: evaluating many thermodynamic quantities, such as the free energy~(\ref{eq free energy}), 
requires knowledge of the HMF~(\ref{eq HMF}). In particular, the partition function 
$\C Z^*_S(\lambda_t) = \C Z_{SB}(\lambda_t)/\C Z_B$ can not be 
infered from the reduced system state~(\ref{eq reduced state}) alone. Instead, it is fixed by the ratio of partition 
functions of the system-bath composite and the bath alone. This is not only theoretically challenging to compute, but 
it also seems experimentally out of reach. 

We here overcome this severe practical limitation in the following sense. First, we show that there is an amount of 
freedom involved in defining the HMF, meaning that the partition function $\C Z^*_S(\lambda_t)$ and therefore the 
thermo\emph{statics} will be different but the thermo\emph{dynamics} remains unchanged. Second, we demonstrate 
that this freedom can be used to construct a strong coupling thermodynamics based solely on local measurements of the 
system. 
Importantly, this is done in a \emph{model-independent} way, based only on three minimal assumptions: the ability 
to measure the system state, knowledge of the system Hamiltonian and knowledge of the bath temperature. 

We start by emphasizing again that the reduced state of $\pi_{SB}(\lambda_t)$, 
\begin{equation}\label{eq HMF eff}
 \pi_S^*(\lambda_t) = \mbox{tr}_B\{\pi_{SB}(\lambda_t)\} 
 = \frac{e^{-\beta\tilde H_S(\lambda_t)}}{\tilde{\C Z}_S(\lambda_t)},
\end{equation}
does not uniquely determine $\tilde H_S(\lambda_t)$ and $\tilde{\C Z}_S(\lambda_t)$. Fixing one, however, determines 
the other. Next, we demonstrate that any choice of $\tilde{\C Z}_S(\lambda_t)$, which fulfills 
\begin{equation}\label{eq freedom}
 \frac{\tilde{\C Z}_S(\lambda_t)}{\tilde{\C Z}_S(\lambda_0)} = \frac{\C Z^*_S(\lambda_t)}{\C Z^*_S(\lambda_0)}
\end{equation}
does not change the thermodynamics. Equivalently, we can say that any choice that fixes the differences of the HMFs, 
i.e., $\Delta H_S^*(\lambda_t) = \Delta\tilde H_S(\lambda_t)$, does not change the thermodynamics. This can be checked 
as follows. First, one expresses the original HMF in terms of the effective HMF from Eq.~(\ref{eq HMF eff}) as 
\begin{equation}\label{eq HMF redefined}
 H_S^*(\lambda_t) = \tilde H_S(\lambda_t) + \frac{1}{\beta}\ln\frac{\tilde{\C Z}_S(\lambda_t)}{\C Z_S^*(\lambda_t)}.
\end{equation}
Notice that the second term on the right hand side is just a real number and can be taken out of any trace 
operation. Using this insight, one readily verifies with the help of Eq.~(\ref{eq freedom}) that 
the thermo\emph{dynamics} (i.e., heat, work, change in internal energy and system entropy, entropy 
production) is insensitive to this redefinition. This is even true for quantities 
defined at the stochastic level. Therefore, we conclude that all choices fulfilled by Eq.~(\ref{eq freedom}) are 
equally legitimate starting points to construct a theory of nonequilibrium thermodynamics. 

Experimentally, reconstructing $\tilde H_S(\lambda_t)$ can be done in various ways, in particular in the classical 
case. For instance, assume that we know the reduced system equilibrium states $\pi_S^*(\lambda_t)$ for all relevant 
values $\lambda_t$ of the control protocol. This state can be infered by doing only measurements of the system. 
Then, set 
\begin{equation}
 \tilde H_S(\lambda_t) = -T[\ln\pi_S^*(\lambda_t) + \ln\tilde{\C Z}_S(\lambda_t)],
\end{equation}
which still does not fully fix $\tilde H_S(\lambda_t)$ as we do not know the constant $\tilde{\C Z}_S(\lambda_t)$. 
However, now we make use of the freedom mentioned above. For this purpose we fix one of the partition functions, 
say the one at time $t=0$, $\tilde{\C Z}_S(\lambda_0)$, to a known value. This value is completely 
arbitrary~\footnote{In our opinion a particularly convenient choice is given by the weak coupling partition function 
$\tilde{\C Z}_S(\lambda_0) = \C Z_S(\lambda_0) = \mbox{tr}_S\{e^{-\beta H_S(\lambda_0)}\}$. Via Eq.~(\ref{eq freedom}) 
this implies $\tilde{\C Z}_S(\lambda_t) = \C Z^*_S(\lambda_t)\C Z_S(\lambda_0)/\C Z^*_S(\lambda_0)$, which guarantees 
that $\tilde{\C Z}_S(\lambda_t)$ always reduces to $\C Z_S(\lambda_t) = \mbox{tr}_S\{e^{-\beta H_S(\lambda_t)}\}$ in 
the weak coupling limit. } 
and fixes $\tilde H_S(\lambda_0)$. To fix $\tilde H_S(\lambda_t)$ for all other times $t\neq 0$, we choose 
$\tilde{\C Z}_S(\lambda_t)$ such that Eq.~(\ref{eq freedom}) is fulfilled, which only requires us to infer 
$\C Z^*_S(\lambda_t)/\C Z^*_S(\lambda_0)$. One way to do so is immediately offered by Eq.~(\ref{eq Jarzynski}) after 
recognizing that $e^{-\beta\Delta\C F_S^*(\lambda_t)} = \C Z^*_S(\lambda_t)/\C Z^*_S(\lambda_0)$. Note that, in 
the classical case, the left hand side of Eq.~(\ref{eq Jarzynski}) can be evaluated by only knowing the stochastic 
work, which can be infered by measuring only system trajectories. 

We comment on another possibility to infer $\tilde H_S(\lambda_t)$ in a \emph{classical} setting provided 
that we fixed $\tilde{\C Z}_S(\lambda_0)$ to an arbitrary value. For this purpose we return to 
Eq.~(\ref{eq EP rel ent}). By using Eqs.~(\ref{eq 2nd law}) and~(\ref{eq free energy}), we see that 
\begin{equation}\label{eq way 1}
 \Delta\lr{H_S^*(\lambda_t)} = W(t) + T\Delta S_\text{Sh}[\rho_S(t)] - T\Sigma(t).
\end{equation}
Here, $\Delta\lr{H_S^*(\lambda_t)} = \mbox{tr}_S\{H_S^*(\lambda_t)\rho_S(t) - H_S^*(\lambda_0)\rho_S(0)\}$ 
denotes the change in expectation value of the HMF and $\Delta S_\text{Sh}[\rho_S(t)]$ denotes the change in 
Shannon entropy of the \emph{classical} distribution $\rho_S(t)$. Now, notice that the right hand side of 
Eq.~(\ref{eq way 1}) is completely determined by knowing $\rho_S(t)$ and $\pi_S^*(\lambda_t)$, but knowledge of the 
HMF is not required to evaluate it. Next, we \emph{use} Eqs.~(\ref{eq freedom}) and~(\ref{eq HMF redefined}) to deduce 
that $\Delta\lr{H_S^*(\lambda_t)} = \Delta\langle\tilde H_S(\lambda_t)\rangle$. Hence, 
\begin{equation}
 \begin{split}\label{eq way 11}
  \mbox{tr}_S\{\tilde H_S(\lambda_t)\rho_S(t)\} =&~ W(t) + T\Delta S_\text{Sh}[\rho_S(t)] \\
  &- T\Sigma(t) + \mbox{tr}_S\{\tilde H_S(\lambda_0)\rho_S(0)\}.
 \end{split}
\end{equation}
Except $\tilde H_S(\lambda_t)$, all quantities are known in this expression and can be infered by measuring the 
system only. To finally reconstruct $\tilde H_S(\lambda_t)$ from this expression, we need a \emph{set} of final states 
$\{\rho_S(t)\}$, which are independent and linearly span the probability space. Such a set can be generated, e.g., 
by using initial states different from $\pi_S^*(\lambda_0)$ (as allowed in the classical 
regime~\cite{SeifertPRL2016}) or by using different driving protocols $\{\lambda_s|0\le s\le t\}$ 
keeping $\lambda_t$ at time $t$ fixed. In contrast to the previously mentioned approach, 
Eq.~(\ref{eq way 11}) might be particularly convenient from a numerical point of view as it only requires 
knowledge about the ensemble of states $\rho_S(t)$. 

We now turn to the quantum case, where the problem is more complicated as the generalization of 
Eq.~(\ref{eq Jarzynski}) can be only derived using the TPPEMS. 
This is experimentally demanding. To circumvent this problem, we consider an adiabatically 
slow process in which $\rho_S(t) = \pi^*_S(\lambda_t)$ for all times $t$ and we assume the 
second law~(\ref{eq 2nd law}) becomes an equality: $W(t) = \Delta\C F_S^*(\lambda_t)$. 
Note that, in contrast to the previous results, the latter is not an exact identity for a finite-size heat bath. 
Instead, we here have to assume that the system-bath composite is coupled to a `superbath' of inverse temperature 
$\beta$. Then, to make the operational meaning of this approach transparent, we express the second law as 
\begin{equation}\label{eq adiabatic}
 \int_0^t ds \mbox{tr}_S\left\{\frac{\partial H_S(\lambda_s)}{\partial s}\pi^*_S(\lambda_s)\right\} 
 = -T\ln\frac{\C Z^*_S(\lambda_t)}{\C Z^*_S(\lambda_0)}.
\end{equation}
This again completely fixes the ratio of partition functions and thus, the HMF up to an irrelevant degree of freedom. 
Note that, in theory, such an adiabatic process requires infinite time. However, compared to the weak coupling regime, 
strong coupling might be helpful here as the relaxation time-scales are larger and hence, we can implement the process 
faster. Furthermore, note that Eq.~(\ref{eq adiabatic}) does \emph{not} require to perform any measurement 
of work \emph{per se}, but is fully accessible by quantum process tomography. 

\section{The criticism of Talkner \& H\"anggi}

In two recent papers~\cite{TalknerHaenggiPRE2016, TalknerHaenggiArXiv2019} Talkner and H\"anggi (abbreviated T\&H in 
the following) critically questioned the approach reviewed in Sec.~\ref{sec intro}. Before turning to their 
three main points of criticism, we review what T\&H take for granted and do not question. In accordance with 
Refs.~\cite{JarzynskiJSM2004, CampisiTalknerHaenggiPRL2009, CampisiTalknerHaenggiJPA2009, HiltEtAlPRE2011, 
PucciEspositoPelitiJSM2013, SeifertPRL2016, StrasbergEtAlNJP2016, PhilbinAndersJPA2016, JarzynskiPRX2017, 
MillerAndersPRE2017, StrasbergEspositoPRE2017, AurellEnt2017, MillerAndersNC2018, PerarnauLlobetEtAlPRL2018, 
AurellPRE2018, HsiangHuEntropy2018, StrasbergEspositoPRE2019, StrasbergPRL2019, HerpichShayanfardEspositoArXiv2019, 
StrasbergQuantum2020} this includes the assumption that the initial state can be taken to be a global Gibbs state 
$\pi_{SB}(\lambda_0)$ and that the average work in both, the quantum and classical case, is given by 
Eq.~(\ref{eq work})~\footnote{In the quantum case, the very meaning of an ``average'' is \emph{a priori} not clear 
as any measurement strategy can disturb the system and change the dynamics. However, if the TPPEMS provides a 
legitimate way to measure work, as assumed by T\&H~\cite{TalknerHaenggiArXiv2019}, then Eq.~(\ref{eq work}) follows 
from an ordinary ensemble average over the TPPEMS statistics.}. They therefore start from the same premise as we did 
in Sec.~\ref{sec intro}. 

Furthermore, T\&H fix the definition of equilibrium energy and entropy by the relations~(\ref{eq equilibrium thermo 
relation}) based on the choice $\C F_S^*(\lambda_t) = -T\ln\C Z_S^*(\lambda_t)$. This choice is referred to as 
``thermodynamically consistent''~\cite{TalknerHaenggiArXiv2019}. But as discussed above, this choice is not unique if 
one only requires the \emph{differences} in thermodynamic state functions to be reproduced and not their absolute value. 
We note that related questions arise on the ongoing discussion of how to correctly account for the interaction energy in 
simple mesoscopic systems~\cite{LudovicoEtAlPRB2014, EspositoOchoaGalperinPRB2015, BruchEtAlPRB2016, 
LudovicoEtAlPRB2016, OchoaBruchNitzanPRB2016, HaughianEspositoSchmidtPRB2018, LudovicoEtAlPRB2018, DouEtAlPRB2018}. 

We now briefly summarize the three main points of criticism by T\&H: 
\begin{itemize}
 \item[A] Since the HMF together with its conventional used normalization [see Eq.~(\ref{eq HMF})] requires precise 
 measurements of $\C Z_S^*(\lambda_t) = \C Z_{SB}(\lambda_t)/\C Z_B$, T\&H ``emphasize that the HMF does not follow 
 from the reduced state of the open system'' and, without additional knowledge, ``the HMF remains undetermined'' and 
 finding it for real systems ``presents in practice an impossible task''~\cite{TalknerHaenggiArXiv2019}. 
 \item[B] When trying to construct the corresponding fluctuating thermodynamic potentials along a single 
 trajectory in view of the classical framework of stochastic thermodynamics, there is a vast amount of ambiguity 
 left. Thus, ``the stochastic energetics suffers from the problem [of ambiguity]'' and ``the same flaw also adheres to 
 stochastic thermodynamics''~\cite{TalknerHaenggiArXiv2019}. Furthermore, T\&H write that ``other restrictions on the 
 hypothetical fluctuating thermodynamic potentials are not known''~\cite{TalknerHaenggiArXiv2019}. 
 \item[C] The Points~A and~B were first put forward in the classical context~\cite{TalknerHaenggiPRE2016}. 
 In addition, in the quantum case T\&H write that ``it is not possible to specify [...] simultaneously work 
 and heat, not even their averages'' and any ``formulation of a first law for other than weakly interacting quantum 
 systems [...] seems doubtful''~\cite{TalknerHaenggiArXiv2019}. 
\end{itemize}
Our reply to this criticism is as follows: 

Concerning Point~A, the main technical contribution of the present paper directly addresses Point~A since we provide 
a clear experimental prescription to determine the HMF, up to a thermodynamically irrelevant constant, by local 
measurements of the system only. This is an important result: although the open system dynamics of $\rho_S(t)$ 
depends strongly on the details of $V_{SB}$ and $H_B$, no knowledge of them is required to experimentally 
infer the thermodynamics of the open system. Thus, the criticism of T\&H expressed in Point~A remains formally 
correct---the HMF \emph{together} with the particular choice~(\ref{eq HMF}) of partition function is \emph{not} 
measurable using only knowledge about $\rho_S(t)$---but this has \emph{no} thermodynamic consequences 
if we choose a partition function obeying Eq.~(\ref{eq freedom}). This less restrictive choice of the partition 
function can be experimentally inferred based only on knowledge about $\rho_S(t)$. 

Concerning Point~B, we first note that the ambiguities in Eqs.~(53) and (54) discussed in Ref.~\cite{TalknerHaenggiPRE2016} are absent if one takes into account that they have to vanish on average for \emph{all} possible time-evolved nonequilibrium states as we show in~\footnote{According to Ref.~\cite{TalknerHaenggiPRE2016}, 
any function $h(\bb x)$ of the system phase space coordinate $\bb x$ can be added to the fluctuating potentials provided 
that its average vanish, $\int d\bb x h(\bb x)p(\bb x;\beta) = 0$. Note that T\&H denote with $p(\bb x;\beta)$ the (in 
general) nonequilibrium system distribution at time $t$ (indicating with the parameter $\beta$ that the bath is 
initially in a conditional equilibrium state with respect to that inverse temperature). The system, however, can be 
prepared and driven arbitrarily, which implies that $\int d\bb x h(\bb x)p(\bb x;\beta) = 0$ must hold for \emph{all 
possible time-evolved system states} $p(\bb x;\beta)$. As those states typically form a complete set of independent 
basis states, it follows that $h(\bb x) = 0$ for each time $t$.}. 
Second, the fluctuating thermodynamic potentials must satisfy the second law~(\ref{eq 2nd law}) or the classical 
fluctuation theorem~(\ref{eq fluctuation theorem}), which indeed constitute further restrictions.

Concerning Point~C, the correct identification of heat and work in the quantum regime is more subtle as 
there is still no consensus on these questions. The main objection of T\&H is based on 
their assessment that heat and work are like ``position and momentum'', whose values ``can not be assigned [simultaneously]'', and the measurements ``need to be error free'' and the ``energy value must be detected 
with certainty''~\cite{TalknerHaenggiArXiv2019}. 
However, such measurements are never strictly realized in any quantum experiment and 
one way to address this issue is to construct a thermodynamic framework that takes into account 
incomplete information, as recently proposed in Refs.~\cite{StrasbergPRE2019, StrasbergWinterPRE2019, 
StrasbergPRL2019, StrasbergQuantum2020}, see also Ref.~\cite{AllahverdyanNieuwenhuizenPRE2005}. This approach 
provides consistent definitions of heat and work 
based on the available information in an experiment and does not assume perfect measurements of the bath 
like the TPPEMS. It also reduces to previously explored cases in the literature in its respective limit.

\section{Conclusion}

The thermodynamic framework based on the HMF provides a solid and, as we have shown, operationally meaningful approach 
to formulate nonequilibrium thermodynamics in the strong coupling regime. It nevertheless has its limitations. Most 
importantly, it does not extend to the experimentally relevant situation of multiple heat baths, where only a few 
formally exact results are known~\cite{EspositoLindenbergVandenBroeckNJP2010, AndrieuxEtAlNJP2009, StrasbergWinterArXiv} 
and a couple of promising theoretical tools, restricted to particular models, were 
devised~\cite{SchallerEtAlNJP2013, EspositoOchoaGalperinPRL2015, EspositoOchoaGalperinPRB2015, 
GelbwaserKlimovskyAspuruGuzikJPCL2015, WangRenCaoSciRep2015, StrasbergEtAlNJP2016, BruchEtAlPRB2016, 
LudovicoEtAlPRB2016, LudovicoEtAlEntropy2016, KatoTanimuraJCP2016, KatzKosloffEnt2016, NewmanMintertNazirPRE2017, 
MuEtAlNJP2017, LudovicoEtAlPRB2018, HaughianEspositoSchmidtPRB2018, SchallerEtAlPRB2018, StrasbergEtAlPRB2018, 
BruchLewenkopfVonOppenPRL2018, WhitneyPRB2018, RestrepoEtAlNJP2018, DouEtAlPRB2018, SchallerNazirBook2018, 
GuarnieriEtAlPRR2019, RestrepoEtAlPRB2019, HisehEtAlJPCC2019, CorreaEtAlJCP2019}. 

To conclude, strong coupling nonequilibrium thermodynamics is not as straightforward as its weak coupling counterpart 
and more care is required when specifying the experimental setup including the different classes of possible system 
preparations. Yet, we are convinced that this quest brought important progress and will continue to do so. 

\subsection*{Acknowledgements}

PS acknowledges fruitful discussions with \'Angel Rivas. Correspondence with Peter H\"anggi and 
Peter Talkner is gratefully acknowledged. PS is financially supported by the DFG (project STR 1505/2-1) and also 
acknowledges funding from the Spanish MINECO FIS2016-80681-P (AEI-FEDER, UE). ME is supported by the European Research 
Council project NanoThermo (ERC-2015-CoG Agreement No. 681456). 


\bibliography{/home/philipp/Documents/references/books,/home/philipp/Documents/references/open_systems,/home/philipp/Documents/references/thermo,/home/philipp/Documents/references/info_thermo,/home/philipp/Documents/references/general_QM,/home/philipp/Documents/references/math_phys,/home/philipp/Documents/references/equilibration}

\appendix
\begin{widetext}
\section{Derivation of the integral fluctuation theorem~(\ref{eq fluctuation theorem})}
\label{sec app}

We denote the phase space coordinates of the system, the bath and the system-bath composite by $x$, $y$ and 
$z = (x,y)$, respectively. The global Hamiltonian reads $H_{SB}(z;\lambda_t) = H_S(x;\lambda_t) + H_B(y) + V_{SB}(z)$. 

We assume that the initial system-bath state is prepared as 
\begin{equation}\label{eq initial state}
 \rho_{SB}(z;0) = \rho_S(x;0)\pi_B(y|x),
\end{equation}
where the conditional state $\pi_B(y|x)$ of the bath is assumed to be equilibrated with respect to $x$: 
\begin{equation}\label{eq conditional bath state}
 \pi_B(y|x) = \frac{e^{-\beta[H_B(y) + V_{SB}(x,y)]}}{\int dy e^{-\beta[H_B(y) + V_{SB}(x,y)]}} 
 = \frac{e^{-\beta[H_{SB}(z;\lambda_0) - H_S^*(x;\lambda_0)]}}{\C Z_B}.
\end{equation}
The last identity follows by using that the HMF in the classical case can be expressed as 
\begin{equation}
 H_S^*(x;\lambda_t) = 
 H_S(x;\lambda_t) - \frac{1}{\beta}\ln\int dy e^{-\beta V_{SB}(x,y)}\frac{e^{-\beta H_B(y)}}{\C Z_B}.
\end{equation}
Note that $\pi_B(y|x)$ does not depend on the value of the control parameter $\lambda_t$. 

We now assume that we generate an ensemble of trajectories by drawing randomly phase space points $z$ sampled from 
the probability distribution~(\ref{eq initial state}). Within the global system-bath state, the time-evolution of a 
point $z$ in phase space is governed by Hamilton's equations of motion. We denote by $z_t = z_t(z_0)$ the time-evolved 
phase space coordinate at time $t$ starting from the initial condition $z_0$. Consequently, $x_t$ and $y_t$ denote the 
projections of $z_t$ on the system and bath phase space, respectively. 

The stochastic work $w$ along a single trajectory $\{z_s(z_0)|s\in[0,t]\}$ is identified as usual with 
\begin{equation}
 w(z_t;t) = \int_0^t ds \dot\lambda_s\frac{\partial H_S(x_s;\lambda_s)}{\partial s} 
 = H_{SB}(z_t;\lambda_t) - H_{SB}(z_0;\lambda_0).
\end{equation}
Furthermore, the stochastic counterpart of the nonequilibrium free energy defined in Eq.~(5) of the main text reads 
\begin{equation}
 f_S^*(x_t;t) = H_S^*(x_t;\lambda_t) + T\ln[h^{N_Sd}\rho_S(x_t;t)]
\end{equation}
Here, $\rho_S(x_t;t)$ is the time-evolved phase space distribution of the system starting from the initial 
condition~(\ref{eq initial state}) evaluated at the system phase space point $x_t$. Furthermore, $h$ is Planck's 
constant, $N_S$ the number of particles in the system and $d$ the dimension (e.g., for a single particle moving 
in one dimension $N_Sd = 1$). This factor was introduced in order to make the argument of the logarithm 
dimensionless. It naturally cancels out when we compute the difference in stochastic free energy, in which we are
actually interested: 
\begin{equation}
 \Delta f_S^*(x_t;t) = f_S^*(x_t;t) - f_S^*(x_0;0) = 
 H_S^*(x_t;\lambda_t) - H_S^*(x_0;\lambda_0) + T\ln\frac{\rho_S(x_t;t)}{\rho_S(x_0;0)}
\end{equation}

We now prove the fluctuation theorem~(\ref{eq fluctuation theorem}) of the main text, whose validity was questioned in 
Ref.~\cite{TalknerHaenggiPRE2016}. For this purpose we first of all note the following useful relation: 
\begin{equation}\label{eq useful identity}
 e^{-\beta[w(z_t;t)-\Delta f_S^*(x_t;t)]} = \frac{\rho_S(x_t;t)\pi_B(y_t|x_t)}{\rho_S(x_0;0)\pi_B(y_0|x_0)}.
\end{equation}
This is a formal mathematical identity, where $\pi_B(y_t|x_t)$ is functionally identical to 
Eq.~(\ref{eq conditional bath state}), but evaluated at the time-evolved phase space coordinate $z_t = (x_t,y_t)$. 
Note that Eq.~(\ref{eq useful identity}) does not assert that the time-evolved conditional state of the bath, defined 
as $\rho_B(y|x;t) = \rho_{SB}(z;t)/\rho_S(x;t)$, is identical to $\pi_B(y_t|x_t)$. Indeed, this is in general not the 
case. 

We can now easily prove the integral fluctuation theorem. First, by definition we have 
\begin{equation}
 \lr{e^{-\beta[w(z_t;t)-\Delta f_S^*(x_t;t)]}} = \int dz_0 \rho_{SB}(z_0;0)e^{-\beta[w(z_t;t)-\Delta f_S(x_t;t)]}.
\end{equation}
Remember that $z_t$ and $x_t$ are functions of the initial phase space point $z_0$. Next, we use 
Eq.~(\ref{eq useful identity}) and afterwards Eq.~(\ref{eq initial state}) to arrive at 
\begin{equation}
 \lr{e^{-\beta[w(z_t;t)-\Delta f_S^*(x_t;t)]}} 
 = \int dz_0 \rho_{SB}(z_0;0) \frac{\rho_S(x_t;t)\pi_B(y_t|x_t)}{\rho_S(x_0;0)\pi_B(y_0|x_0)}
 = \int dz_0 \rho_S(x_t;t)\pi_B(y_t|x_t).
\end{equation}
Now, we perform a change of variables $z_0\rightarrow z_t$, which---by virtue of Liouville's theorem---finally yields 
\begin{equation}
 \lr{e^{-\beta[w(z_t;t)-\Delta f_S^*(x_t;t)]}} = \int dz_t \rho_S(x_t;t)\pi_B(y_t|x_t) = 1.
\end{equation}
\end{widetext}

\end{document}